# Adiabatic Control of Spin-Wave Propagation using Magnetisation Gradients

Marc Vogel[1,*], Rick Aßmann[1], Philipp Pirro[1], Andrii V. Chumak[1], Burkard Hillebrands[1] & Georg von Freymann[1,2]

[1] Department of Physics and State Research Center OPTIMAS, University of Kaiserslautern, Erwin-Schroedinger-Str. 56, 67663 Kaiserslautern, Germany

[2] Fraunhofer-Institute for Industrial Mathematics ITWM, Fraunhofer-Platz 1, 67663 Kaiserslautern, Germany

[*] Correspondence and requests for materials should be addressed to M. Vogel (email: mvogel@physik.uni-kl.de)

**Abstract**

Spin waves are of large interest as data carriers for future logic devices. However, due to the strong anisotropic dispersion relation of dipolar spin-waves in in-plane magnetised films the realisation of two-dimensional information transport remains a challenge. Bending of the energy flow is prohibited since energy and momentum of spin waves cannot be conserved while changing the direction of wave propagation. Thus, non-linear or non-stationary mechanisms are usually employed. Here, we propose to use reconfigurable laser-induced magnetisation gradients to break the system's translational symmetry. The resulting changes in the magnetisation shift the dispersion relations locally and allow for operating with different spin-wave modes at the same frequency. Spin-wave momentum is first transformed via refraction at the edge of the magnetisation gradient region and then adiabatically modified inside it. Along these lines the spin-wave propagation direction can be controlled in a broad frequency range with high efficiency.

**Introduction**

In modern electronics, three-dimensional complex metal circuits are the backbone to transport information between data processing units. To overcome the limitations inherent to electronics, e.g. limited bandwidth and Ohmic losses to name a few, spin waves, and their quanta magnons, are considered to be the information carriers for next generation data-processing devices [1-6]. In-plane magnetised films are preferred due to the minimised demagnetisation energy. For this configuration, important building blocks for magnonic circuits have already been realised, e. g.

magnon transistors [7], majority gates [8,9] or multiplexers [10]. However, the dispersion relations of dipolar magnetostatic and dipole-exchange spin waves, which are usually used for data processing, are strongly anisotropic. This anisotropy has to be taken into account for efficient data transfer between the building blocks.

For dipolar spin-waves in in-plane magnetised films two modes can be distinguished: backward volume magnetostatic spin waves (BVMSWs) propagating along the static magnetisation and magnetostatic surface spin waves (MSSWs) propagating perpendicularly, respectively. For a fixed saturation magnetisation and a fixed external magnetic field, the frequencies of MSSWs lay above the ferromagnetic resonance (FMR) frequency whereas the frequencies of BVMSWs lay below. Hence, a change of the propagation direction with respect to the magnetisation requires mode conversion between MSSWs and BVMSWs, which energy conservation obviously prohibits in homogeneous films (as opposite to the out-of-plane magnetisation case [11]).

To overcome this problem two roads can be followed: One can adjust the frequency gap via non-linear and non-stationary mechanisms, e. g. multi-magnon scattering processes [12-14], dynamic systems like dynamic photonic / magnonic crystals [15,16], multi-particle scattering processes [17,18] (e. g. Brillouin or Raman scattering), and the Doppler effect [19-21]. Or, one can change the spin-wave propagation direction while conserving the spin-wave frequency via non-uniform magnetic media (saturation magnetisation and/or biasing magnetic field [10,22-24]) or lateral confinement of the magnetic structures [25,26]. In the latter case, the locally varying magnetic parameters shift the spin-wave dispersion relations and, hence, allow for the conversion between different spin-wave modes. However, the usual dimensions $\delta x$ of such inhomogeneities are small in order to provide a large change in the spin-wave momentum $|\vec{k}_{\text{inh}}| \propto 1/\delta x$. The underlying elastic two-magnon scattering process [27] $\vec{k}_{\text{new}} = \vec{k}_{\text{old}} + \vec{k}_{\text{inh}}$ effectively changes the propagation direction. Such high localisation of the inhomogeneities is associated with undesirable spin-wave reflections and offers no flexibility in spin-wave steering.

Here, we present an adiabatic and reflection-less approach to change the spin-wave propagation using graded index spin-wave media extending over sizes $\delta x$ much larger than the spin-wave wavelength $\lambda = 2\pi/|\vec{k}| \ll \delta x$. A continuous transformation of spin-wave momentum in magnetisation gradients leads to highly-efficient mode conversion over a wide frequency range. We demonstrate the control of spin-wave propagation by optically reconfiguring the media to tune the transformation of spin-wave momentum.

**Results**

**Conversion principle.** Let us discuss the mechanism behind the two-dimensional magnon guiding for the case of converting a pure BVMSW into a pure MSSW, propagating perpendicularly to each other. Fig. 1a depicts this configuration. The pure initial and final modes propagate in regions of the magnetic waveguide with saturation magnetisation $M_{S,1}$ (blue region) and $M_{S,2}$ (red region), respectively. To obey energy conservation, the dispersion relations of the two modes have to overlap, requiring $M_{S,2} < M_{S,1}$ (see solid blue line and dashed red line in Fig. 2b, overlap is found in the grey marked conversion area). To additionally fulfil momentum conservation, the wavevectors adiabatically have to be transformed into each other. This adiabatic transformation is enabled by a gradient of the saturation magnetisation perpendicular to the initial propagation direction (grey bordered area in Fig. 1a) and spatially represents the area in which the dispersion relations overlap. To illustrate the flow of spin-wave energy during the conversion process, the local spin-wave group velocities $\vec{v}_{gr}$ are depicted in Fig. 1a by black arrows. This process is initiated by refraction of the spin wave [28] at the edge of the magnetisation gradient (orange line in Fig. 1a, modified wavelength but no change in the direction of $\vec{k}$ for our simple example but important later on) and continues with quasi-adiabatic transformation of the spin-wave wavevector [29] (highlighted with coloured arrows at certain positions). This scheme also holds for more general cases in which spin waves propagate under arbitrary angles $\varphi = \angle(\vec{k}, \vec{H}_{ext})$ with respect to the biasing magnetic field $\vec{H}_{ext}$.

We realise the different saturation magnetisations for an experimental demonstration of this mode-conversion process by laser-induced local heating recently introduced by us [30]. This method allows for generating almost arbitrary saturation magnetisation landscapes (see methods) to influence the spin-wave propagation. For the simple example discussed above (Fig. 1a) the BVMSW is launched in a cold region (blue) and converted into the MSSW in a hot region (red). For launching and detecting the respective spin waves, we use the microwave antenna configuration depicted in Fig. 1c.

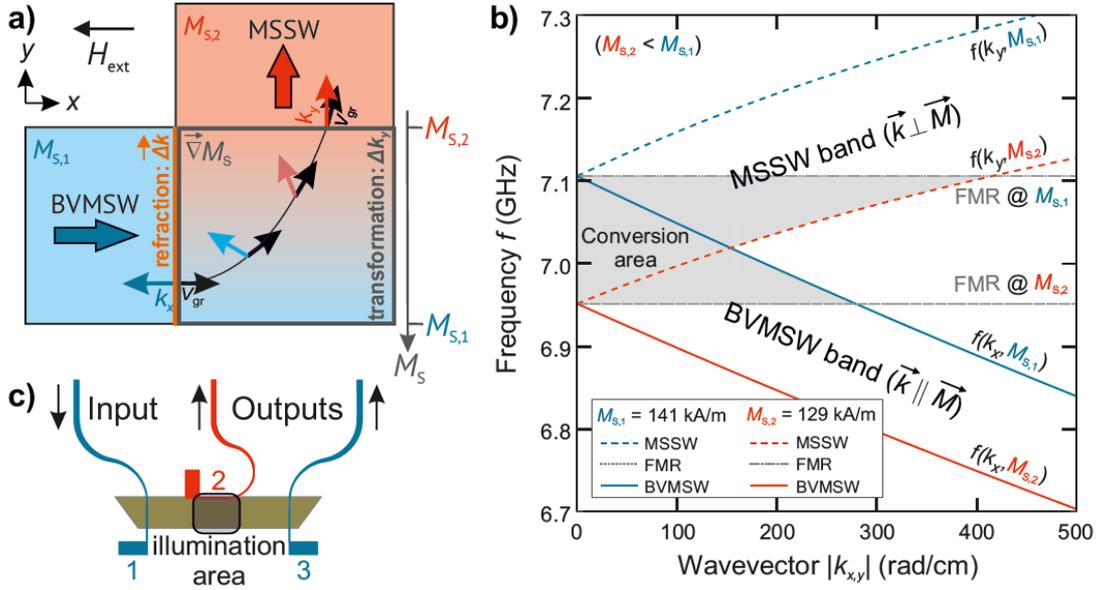

**Figure 1 | Scheme of the mode conversion and dispersion relations for different local values of the saturation magnetisation. a)** The mode conversion process transforms BVMSW (blue) to MSSW (red) modes. Both types of spin waves propagate in the sketched areas at different values of the saturation magnetisation $M_S$. A magnetisation landscape – a temperature-induced arbitrary shaped magnetisation gradient $\vec{\nabla} M_S(x,y)$ – enables the mode conversion. A two–step process changes the direction of the wavevector $\vec{k}$ and the group velocity $\vec{v}_{gr}$. First, the spin-wave is refracted at the interface to the gradient decreasing the wavenumber. Next, a quasi-adiabatic transformation inside the magnetisation landscape occurs. The external magnetic field $\vec{H}_{ext}$ is oriented parallel to the waveguide (in-plane orientation). **b)** The dispersion relations of MSSW (propagating in the $y$ direction) and BVMSW modes (propagating in the $x$ direction) are shown versus the wavevector $k$. Due to the saturation magnetisation gradient, a wide band for both modes is realised. The intersection of these bands forms the conversion area (grey), where the mode conversion is allowed. The values $M_{S,1}$ and $M_{S,2}$ correspond to the temperatures $T_1$ at antenna 1 and $T_2$ at antenna 2 and are chosen according to the infrared camera measurements in Fig. 2. The conversion area is limited by the ferromagnetic resonance frequencies (FMR) for $M_{S,1}$ and $M_{S,2}$, respectively. **c)** The experimental set-up is shown schematically. A light pattern which creates a magnetisation gradient is formed close to the centre in $x$-direction of the waveguide at antenna 2 (illumination area). With respect to the externally applied magnetic field $\vec{H}_{ext}$, antennas 1 and 3 can excite and detect BVMSWs. Antenna 2 is used for the detection of MSSWs. Thus, spin waves excited at $M_{S,1}$ are converted and afterwards measured at $M_{S,2}$.

**Experimental studies of mode conversion in magnetisation gradients.** We perform our experiments on a 6.6 µm thick ferrimagnetic Yttrium Iron Garnet (YIG) film acting as spin-wave waveguide. The dependency of its saturation magnetisation $M_S$ on the local temperature $T(x,y)$ is well known: $M_S(T)$ decreases for increasing $T$ [29-32] (see equation 3 in the methods part). To heat the magnon waveguide, arbitrary intensity distributions are realised via computer-generated holograms. Green laser light from a continuous-wave laser ($\lambda$=532 nm) illuminates a phase-only spatial light modulator on which the computer-generated holograms are displayed, resulting in a controlled intensity distribution on the spin-wave waveguide. As the optical set-up is similar to the one presented in [30], we refer to the methods part and supplementary materials for further details. The YIG film absorbs approximately 55% of the green light. Remaining light impinges onto a black absorber layer which further increases the thermal contrast inside the spin-wave waveguide [30]. An infrared camera records the resulting temperature distribution over the sample. The corresponding local saturation magnetisation is calculated using the given spatial temperature distribution (see methods). To excite and detect spin waves, a vector network analyser is connected to the three-microstrip-antenna configuration shown in Fig. 1c to determine the scattering parameters $S_{21}$ and $S_{31}$.

Let us discuss now the experimental results for the exemplary cases shown in Fig. 2. Figure 2a depicts transmission spectra for spin-wave transmission from antenna 1 to 3 (parameter $S_{31}$). Without any intensity distribution, $S_{31}$ follows a typical BVMSW transmission spectrum [33] with a maximum of about -24 dB. This value is defined by the excitation/detection efficiency of the antennas and by the damping of spin waves propagating over 10 mm distance. The grey shaded area (7.10–7.15 GHz) marks the spectral region above the ferromagnetic resonance frequency for an unheated sample. In this frequency range, only higher width-modes (pronounced peak) can be excited by antenna 1. However, spin-wave modes with frequencies above 7.1 GHz are not of interest in this work. The picture dramatically changes, if we shine either one of the two exemplary intensity distributions onto the sample: for a rectangular (solid blue curve) as well as for a triangular distribution (solid red curve) the transmission from antenna 1 to antenna 3 vanishes completely.

In Fig. 2b, we observe the opposite behaviour. Transmission from antenna 1 to 2 (parameter $S_{21}$) for the reference measurement (black solid line) shows practically no signal: The constant background around -50 dB over a wide frequency range up to 7.1 GHz is due to the electromagnetic leakage between the antennas. The absence of the $S_{21}$ transmission shows that no excitation of waves with $\vec{k} \perp \vec{H}_{\text{ext}}$ occurred and, consequently, no BVMSW-to-MSSW mode conversion is possible, as expected. For the rectangular (blue solid line) and triangular (red solid lines) pattern, a strong transmission signal is observed, comparable in strength to the one measured for the BVMSW in the reference measurement (Fig. 2a). This behaviour is consistent with mode-conversion from BVMSW to MSSW modes. To explain this further let us discuss the resulting saturation magnetisation gradient in more detail. Both light patterns are intensity gradients reaching from the maximum intensity (upper waveguide edge at antenna 2) down to 20% of this value (lower waveguide edge, see bottom part of Fig. 2c). Due to the intrinsic thermal conductivity of YIG a temperature gradient $\vec{\nabla}T(x,y)$ and, thus, a magnetisation gradient $\vec{\nabla}M_S(x,y)$ evolves inside the sample. We keep the saturation magnetisation $M_{S,2}(T)$ at antenna 2 approximately constant by adjusting the laser power appropriately. This is done to allow a quantitative comparison of the two cases. Antenna 1 and antenna 3 are kept far away from the heated area, thus $M_{S,1}$ and $M_{S,3}$ are almost equal and correspond approximately to the value at room temperature. By non-uniformly heating the sample, the $S_{21}$-parameter maximum increases up to -23 dB for the rectangular pattern and -22 dB for the triangular pattern. Compared to the maximum of the BVMSW spectrum $S_{31}$ (-24 dB), the mode conversion in magnetisation gradients is very effective. Furthermore, the bandwidth of the converted spectrum at -30 dB increases from 38 MHz in the rectangular case to 84 MHz in the triangular case. In contrast to the unheated sample, now the $S_{31}$ parameter shows practically no transmission since the spin waves are converted and partially reflected in the heated region.

The temperature profiles – recorded using an infrared camera – are used to calculate the local saturation magnetisation in dependency of the $y$ coordinate (Fig. 2c) at the centre of the illumination area (grey dotted lines). To allow for an effective mode conversion the $M_S$ gradient needs a negative slope (away from antenna 2) and has to reach all over the waveguide width. This is consistent with the band structure model described above: MSSW with the same frequency as BVMSW can propagate only in a region of YIG with decreased $M_S$. So, the crucial requirement for the conversion is that antenna 2 should detect spin waves from a part of the waveguide at lower saturation magnetisation compared to antenna 1. Moreover, smooth and linear magnetisation transitions are needed to avoid reflections [32]. At the lower ($y$=0.0 mm) and upper ($y$=3.0 mm) edges of the waveguide von Neumann boundary conditions reduce the temperature and, thus, increase the saturation magnetisation. Moreover, antenna 2 acts as a heat sink.

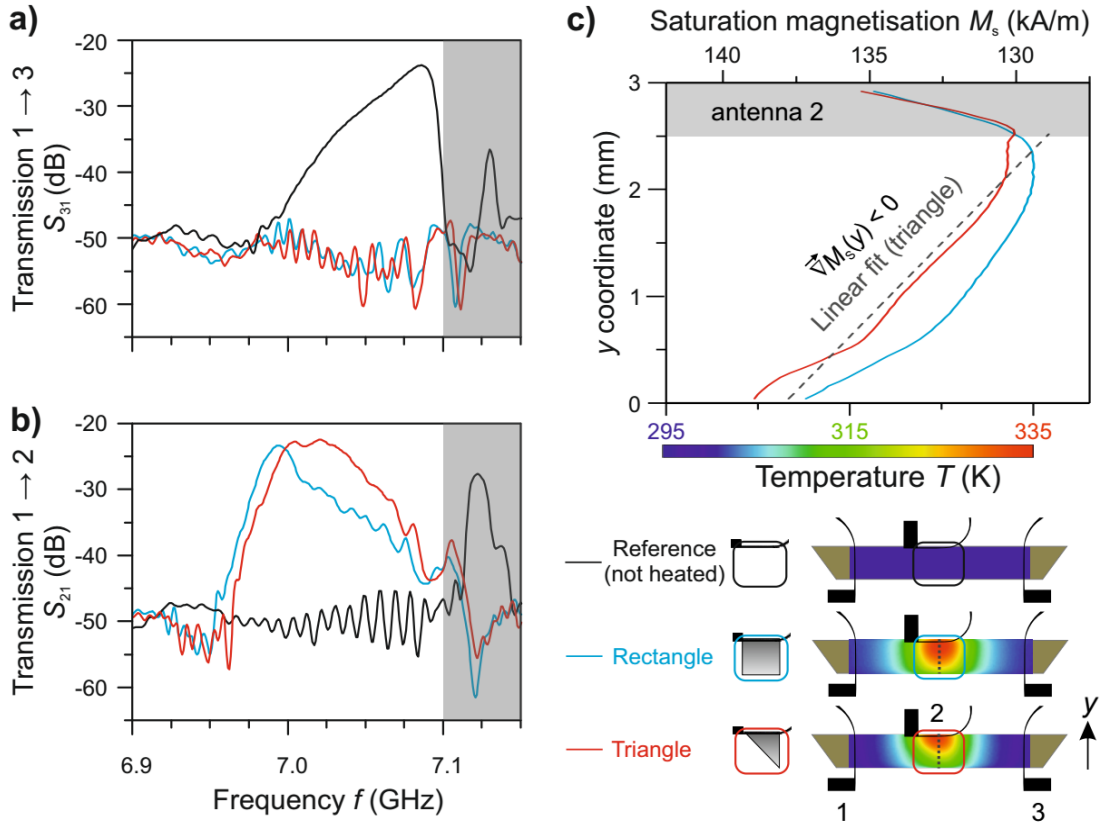

**Figure 2 | Transmission characteristics for different light intensity distributions and the corresponding local saturation magnetisation.** A rectangular and triangular graded intensity distribution is applied in the centre between antennas 1 and 3 – at the position of antenna 2 ("illumination area" in Fig. 1c) – and the corresponding spin-wave transmission $S_{31}$ **(a)** and $S_{21}$ **(b)** is measured (rectangle: blue line, triangle: red line) and compared with the reference measurement without heating (black line). The light-induced temperature distributions are shown as colour code (bottom right corner of Fig. 2, blue: cold, red: hot). The $S_{31}$ ($S_{21}$) parameter depicts the transmission of spin waves form antenna 1 to 3 (1 to 2). The grey area (7.10 – 7.15 GHz) is above the ferromagnetic resonance frequency for an unheated sample (see the $S_{31}$ parameter) and is not of interest in this work. The strongest conversion (up to -22 dB) is achieved by a triangular light pattern with an intensity gradient. The small frequency shift between the triangle's and the rectangle's spectrum in Fig. 2b of (11±1) MHz, exemplarily measured at a -40 dB level, is due to an incomplete match to the temperatures near antenna 2. **c)** The profiles of the saturation magnetisation along the $y$ coordinate (over the waveguide width) at the centre of the illumination area (grey dotted line) are shown for the different light distributions. Gradients in the saturation magnetisation with a negative slope cause the mode conversion. The slope of the linear fit (from 0.0 to 2.75 mm) for the triangle with an intensity gradient is -3.58 kA/m per mm.

**Conversion area and efficiency maximum.** To experimentally check the limits of the conversion area shown in Fig. 1b, the sample is illuminated with the triangular light distribution with an intensity gradient (see Fig. 2). With increasing laser power, the spin-wave waveguide heats up. Accordingly, $M_S$ decreases at antenna 2 and, thus, the difference between the saturation magnetisations $\Delta M_S$ at antenna 1 and 2 increases. $\Delta M_S$ depicts qualitatively the gradient of the saturation magnetisation. Figures 3a and 3b show the corresponding transmission spectra as a colour map for the $S_{21}$ and $S_{31}$ parameters, respectively. Stronger heating and, therefore, reduction of $M_S(T)$ leads to a wider frequency bandwidth of the conversion area, as expected from our reasoning above (the conversion area in Fig. 1b depends on the difference in the saturation magnetisations $M_{S,1}$ and $M_{S,2}$). Spin waves cannot propagate to antenna 3 by increasing the $M_S$ gradient at antenna 2 (Fig. 3b).

Furthermore, we calculate the conversion efficiency qualitatively (Fig. 3c) using the $S_{21}$ parameter, the frequency range of the antenna excitation and detection, and the propagation parameters of the spin waves (see methods). The corresponding frequency of the efficiency maximum is determined via numerical peak fitting. The conversion takes place in between the ferromagnetic resonance frequencies at antenna 1 and 2 (FMR$_1$ and FMR$_2$). The frequency position of the efficiency maxima can be compared to the dispersion relations of the BVMSW and MSSW modes (Fig. 3d). The strongest conversion is observed at the crossing point of the dispersion relation of the MSSW mode at antenna 2 and the dispersion relation of the BVMSW at antenna 1 (see Fig. 1b). For this particular frequency, the magnitude of the momentum is conserved.

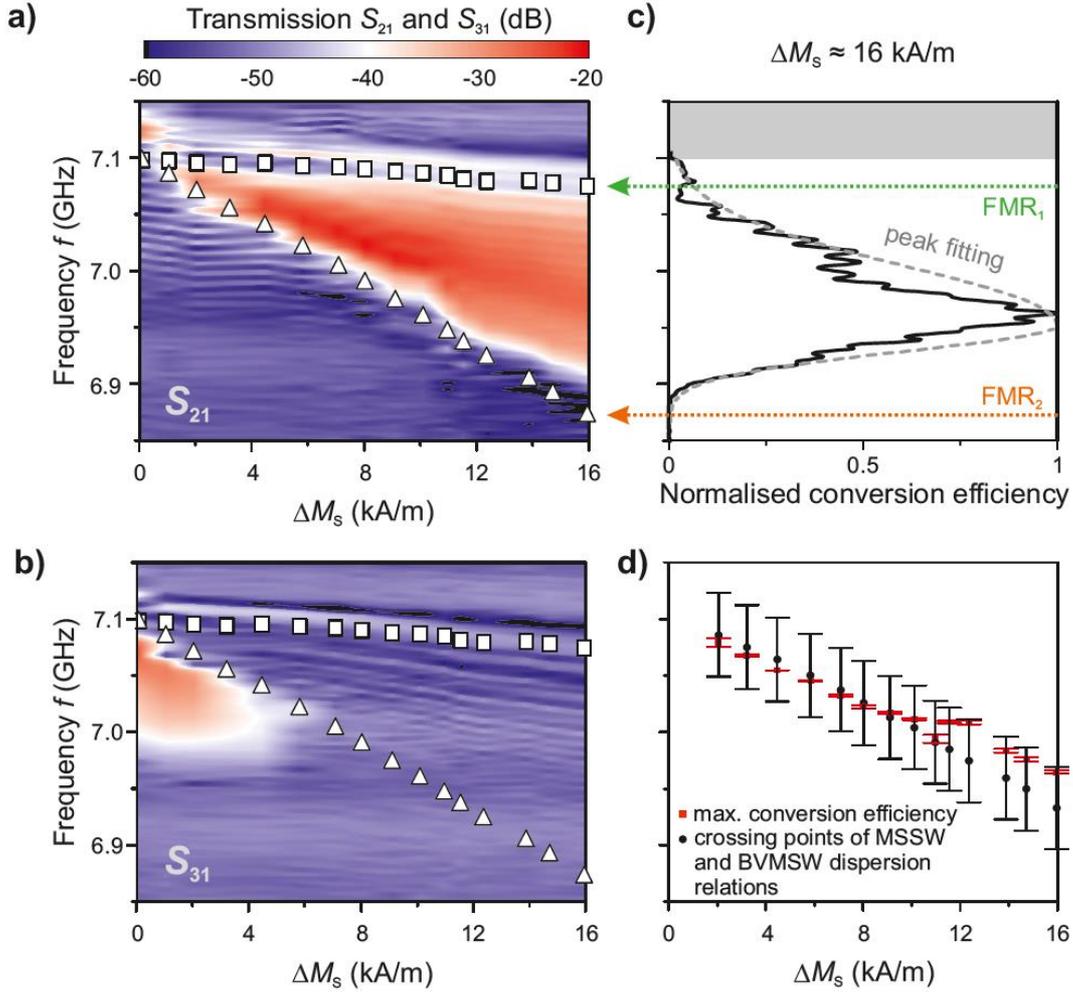

**Figure 3 | Experimental magnetisation dependence of the conversion area and the conversion efficiency's maximum.** The frequency $f$ is shown versus the difference between the saturation magnetisations $\Delta M_S = M_{S,1} - M_{S,2}$ at antenna 1 and 2 for the $S_{21}$ **(a)** and $S_{31}$ **(b)** parameters, respectively. The spin-wave transmission is colour coded (red: strong transmission, blue: no transmission). White squares (triangles) are the experimentally determined FMR frequencies using the reflection parameters $S_{11}$ ($S_{22}$) at antenna 1 (antenna 2). The width of the conversion area increases with decreasing saturation magnetisation (a) and the FMR frequency at antenna 2 shifts to lower frequencies. On the other hand, spin waves with a frequency larger than this FMR frequency cannot travel through the magnetisation landscape to reach antenna 3 (b). **c)** The normalised efficiency of the mode conversion (black line) of BVMSWs to MSSWs is shown versus the frequency $f$. Here, the saturation magnetisation difference $\Delta M_S$ is about 16 kA/m. The position of the maximum is determined via fitting the experimental data with an empirical bi-Gaussian function (grey dotted line). **d)** The frequency of the numerical conversion efficiency peak (red squares) is plotted versus $\Delta M_S$. Black circles indicate the theoretically expected position of the crossing points of the BVMSW (at antenna 1) and MSSW (at antenna 2) dispersion relations for the corresponding saturation magnetisations. The error

bars result from the peak fitting error of the experimental data and the error in the external magnetic field ($\mu_0 \Delta H_{\text{ext}} = \pm 1\,\text{mT}$), respectively.

**Isofrequency curves and micromagnetic simulations.** In order to obtain deeper insight into the conversion mechanism, we perform numerical micromagnetic calculations for magnetisation variations reaching from $M_{S,1} = 141\,\text{kA/m}$ (at room temperature) to $M_{S,2} = 123\,\text{kA/m}$ ($T_2 \approx 353\,\text{K}$, largest temperature investigated in the experiments, for details of the calculations see the methods section). If the spin waves are excited at $f=7.0\,\text{GHz}$ the mode conversion is energetically allowed (see dispersion relations in the supplementary materials). As a reference, additional simulations for 6.8 GHz are shown in the supplementary materials (pure BVMSW behaviour). The propagation of spin waves in gradients of the saturation magnetisation is visualised in Fig. 4 and can be understood via the according isofrequency curves [34,35]. In general, mode conversion can occur if the angle between wavevector and external magnetic field $\varphi = \angle(\vec{k}, \vec{H}_{\text{ext}})$ is larger than a critical angle $\varphi_c$ [36]. MSSWs are described by isofrequency curves, which lie completely above the line corresponding to $\varphi_c$ ($\varphi_c \approx 46°$, see supplementary materials). Curves below or even crossing this limit are BVMSW modes propagating under the angle $\varphi$. We identify the two mechanisms changing the direction of the wave propagation: first, refraction at the interface or at the shape of the magnetisation gradient [28] and, second, the gradient $\vec{\nabla} M_S(x,y)$ itself [29]. For a rectangular magnetisation landscape (Fig. 4a), the waveguide is divided into two regions: one with saturation magnetisation below $M_{S,\text{FMR}}$ and one with a value above. As example, we choose $M_{S,\text{FMR}}$ (green dotted line) as the value of the saturation magnetisation at which the FMR frequency equals 7.0 GHz (excitation frequency in the micromagnetic simulations). In this case, the propagating spin waves (black dotted line) at $M_S \approx 141\,\text{kA/m}$ enter the gradient at a certain value $M_{S,\text{in}}$ defined by the coordinate in the direction perpendicular to the propagation direction. The spin waves cannot enter the gradient region where $M_{S,\text{in}} < M_{S,\text{FMR}}$ (see region above green dotted line) and, thus, they are reflected. For $M_{S,\text{in}} \geq M_{S,\text{FMR}}$ the wave propagates into the magnetisation gradient and only its wavelength is modified (no change in the direction of $\vec{k}$). Afterwards, mode conversion takes place: the spin-wave wavevector and, respectively, its wavelength change differently in different positions over the waveguide's width resulting in a bending of the phase fronts and in an adjustment of the group velocity's direction. The spin waves propagate into regions with decreased $M_S$ adiabatically. The corresponding isofrequency curves illustrate how this process is happening: the spin waves at $M_S \approx 141\,\text{kA/m}$ propagate into the gradient region at $M_{S,\text{in}}$. The tangential component of $\vec{k}$ (with respect to the interface) is conserved and only $k_x$ changes. After entering $\vec{\nabla} M_S(x,y)$, $k_x$ is conserved and only $k_y$ is modified since the translational symmetry of the system is broken due to the magnetisation gradient in the $y$ direction. The change $\Delta k_y$ is due to quasi-adiabatic refraction of the spin waves in the gradient area, whose extend is much larger than the wavelength [29]. Eventually, $\vec{k}$ rotates in the $x$-$y$-plane.

A similar behaviour is observed in the triangular case (Fig. 4b) for $M_{S,\text{in}} \geq M_{S,\text{FMR}}$. The spin wave is converted adiabatically while propagating towards the region of lower saturation magnetisation. For the triangle, additional refraction [28] due to the shape of the $\vec{\nabla} M_S(x,y)$ area occurs. The wave propagates towards the interface under an angle of 45° and, thus, $k_x$ and $k_y$ are modified. Next, $\vec{\nabla} M_S(x,y)$ leads to an increase in $k_y$. The triangular case (Fig. 4b) for $f$=7.0 GHz shows minor reflections at values of the magnetisation in the range of $M_{S,\text{FMR}} \leq M_S \leq 124\,\text{kA/m}$. In the $\vec{k}$-space, the black dotted line corresponding to the refraction now crosses the MSSW isofrequency curves (above the grey dotted line associated to $\varphi_c$) and a direct conversion from BVMSWs to MSSWs due to the geometrical shape is possible. The gradient of $M_S$ is used afterwards to tune the propagation direction of MSSWs to the demanded angle that is perpendicular to the spin-wave waveguide. Thus, the efficiency of the mode conversion is improved. This is in direct agreement with our experimental results. All the simulations show reflections at the top of the waveguide due to the change in the $y$-component of the group velocity. For more clarity, further calculations (no upper waveguide edge) have been performed and can be found in the supplementary materials. Also, additional movies are presented in the supplementary materials, showing the excitation of BVMSWs for different frequencies and the spin-wave propagation into and through the rectangular and triangular magnetisation landscape, respectively.

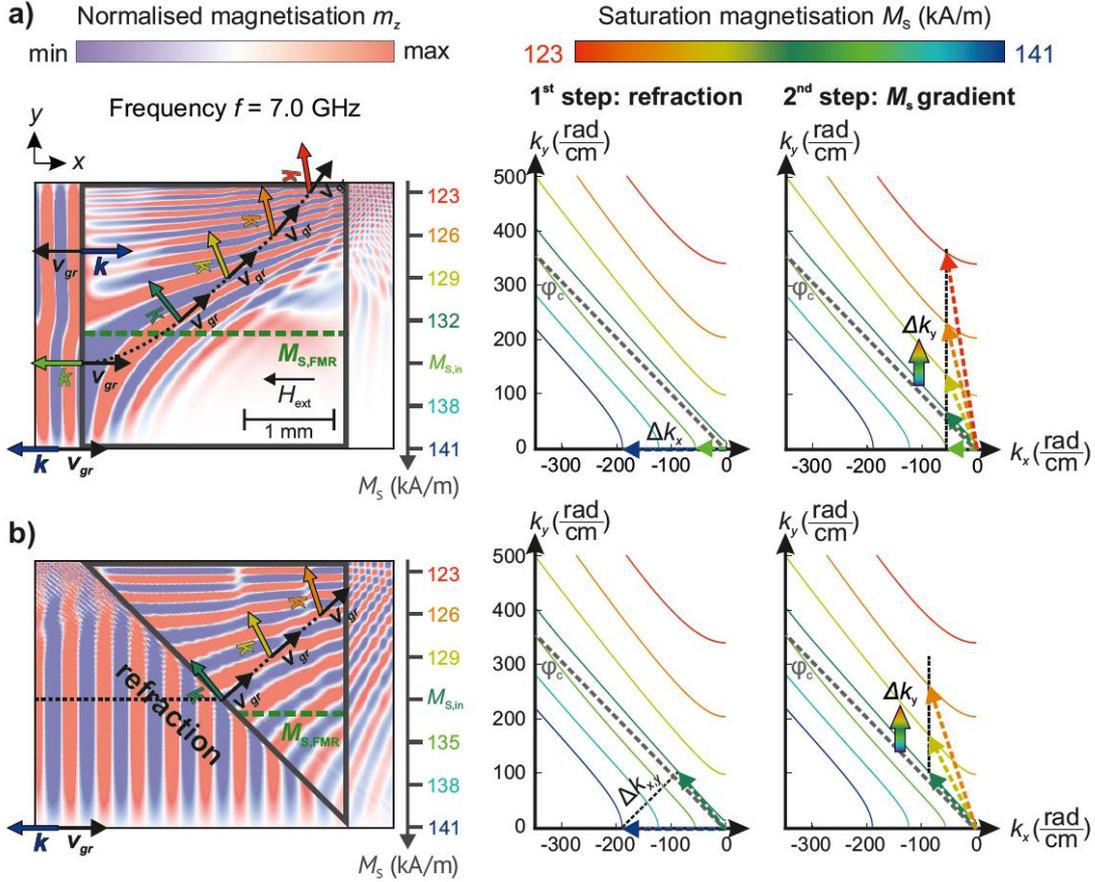

**Figure 4 | Micromagnetic calculations (left) and corresponding isofrequency curves (right).** The numerical calculations are performed for $f = 7.0$ GHz. Two gradient shapes are analysed: a rectangle **(a)** and a triangle **(b)**. Fig. 4a is shown schematically in Fig. 1a. The isofrequency curves illustrate how the wavevector $\vec{k}$ is transformed while the spin wave propagates (dotted line) into and through the magnetisation gradient region. The spin wave enters the gradient at a certain local value of the saturation magnetisation $M_{S,in}$. The isofrequency curves visualise the 2-step conversion process: first, the spin waves are refracted at the interface. Next, the transformation inside the gradient takes place. The direction of $\vec{k}$ can be directly compared to the micromagnetic simulations. Here, the spin wave is excited at the left of each picture and propagates to the right into the magnetisation gradient region (grey rectangle or triangle) where the conversion can take place. The dynamic magnetisation $m_z$ is colour coded (red: max, blue: min).

**Conclusion.** We proposed and demonstrated experimentally as well as numerically mode conversion between backward volume and magnetostatic surface spin-wave modes using gradients of the saturation magnetisation $M_S$. The magnetisation landscapes are achieved via reconfigurable laser-induced thermal patterns. The gradients perpendicular to the spin-wave propagation direction allow for the 90 degree rotation of the spin-wave wavevector via breaking the translational symmetry of the waveguide. As opposite to the spin-wave momentum, the energy of spin waves is conserved in this process due to the intersection of frequency bands of both spin-wave modes allowed in a non-uniform media. Since the proposed conversion is an adiabatic process (the characteristic length scales of the gradients are much larger than the spin-wave wavelength) it allows for the minimisation of undesirable reflections and ensures high efficiency of the conversion in a wide range of spin-wave frequencies. Moreover, it is shown that the conversion mechanism can be further enhanced via the exploiting of refraction at the edge of the magnetisation gradient. Thus, the mode conversion using a triangle-shaped area of the magnetisation gradient has shown even higher efficiency in a wider frequency range. The concept of spin-wave guiding in in-plane magnetised films using magnetisation gradients can be extended to the nanoscale (using e.g. ionic implantation [37-39]) and is promising for the realisation of magnonic networks for novel computing concepts. Moreover, the proposed approach is applicable for the guiding of waves in strongly-anisotropic media independently on the nature of these waves.

## Methods

**Sample.** The $W$=3 mm wide sample consists of a multilayer structure. The substrate is a 500 µm thick Gadolinium Gallium Garnet (GGG, paramagnetic) layer. On top, a 6.6 µm thin Yttrium Iron Garnet (YIG, ferrimagnetic) film was deposited via liquid phase epitaxy. A several 10 µm thick black lacquer is used as additional absorber of the light intensity to increase the thermal contrast of the temperature profile – due to an approximately one order of magnitude smaller thermal conductivity compared to YIG [30]. Furthermore, the lacquer is used as glue. The dielectric spacer is an adhesive tape with a thickness of about 50 µm. This spacer separates the antennas – otherwise they would act as a heat sink – thermally from the sample. GGG is close-to-transparent for green light (532 nm wavelength). Most of the absorption takes place in the YIG film (approx. 55%). Remaining light impinges onto the black absorber. A scheme of the multilayer system is shown in the supplementary materials.

**Experimental setup.** The experimental setup is divided into two parts – optics and microwave technology. The illumination of the sample is realised using a green continuous wave laser (Coherent Verdi G7, 532 nm, maximal power of 7.4 W). The beam is expanded / confined via a combination of a Galilean and a Keplerian telescope. Finally, a Fourier lens reconstructs the hologram on the sample. About thirty-two percent (32%) of the input laser power is available for the heating process ($P_{\text{holo}} = 0.32 \, P_{\text{laser}}$). The intensity distributions on the sample are achieved by a spatial light modulator (Hamamatsu X10468-01; phase only modulation). Multi foci [40] are used to realise the applied rectangular and triangular intensity distributions. The intensity gradients used in the experiment range from the highest intensity down to 20 % of this value. The resulting temperature distributions are recorded using an infrared camera (FLIR SC-655). A scheme of the optical set-up can be found in the supplementary materials.

The microwave excitation and detection is performed by a vector network analyser (Anritsu MS4642B). To avoid nonlinearities, the microwave power for the spin-wave excitation was set to -5 dBm. The copper antennas are about 18 µm thick and 50 µm wide along the excitation/detection area of 3 mm (fabricated on a Duroid ceramic substrate). The distance $L$ between antenna 1 and 3 is 10 mm. The centre of antenna 2 is located in between (5 mm distance to the other antennas). The fabrication uses standard lithography and etching processes. The external magnetic field $H_{\text{ext}}$ is created via an electromagnet (Lakeshore EM4-HVA, $H_{\text{ext,max}} \approx 400$ mT @ 110 mm pole distance).

**Dispersion relations.** To discuss the spin-wave mode conversion analytically, we use the respective dispersion relations for the BVMSW and MSSW modes:

$$f_{\text{BVMSW}}(k_x, T) = \frac{1}{2\pi} \sqrt{\omega_{\text{H}} \left[ \omega_{\text{H}} + \omega_{\text{M}}(T) \cdot \left( \frac{1 - e^{-k_x d}}{k_x d} \right) \right]} \qquad (1)$$

$$f_{\text{MSSW}}(k_y, T) = \frac{1}{2\pi} \sqrt{\left( \omega_{\text{H}} + \frac{\omega_{\text{M}}(T)}{2} \right)^2 - \left( \frac{\omega_{\text{M}}(T)}{2} \right)^2 \cdot e^{-2 k_y d}} \qquad (2)$$

Where $\omega_{\text{H}} = \gamma \mu_0 H_{\text{ext}}$, $\omega_{\text{M}}(T) = \gamma \mu_0 M_{\text{S}}(T)$, and $d = 6.6\,\mu\text{m}$ is the YIG film thickness. Here, $\gamma = 2\pi \cdot 28.0\,\text{GHz/T}$ is the electron gyromagnetic ratio and $\mu_0 = 4\pi \cdot 10^{-7}\,\text{N/A}^2$ the permeability in vacuum. $\mu_0 H_{\text{ext}} = (180 \pm 1)\,\text{mT}$ is the external applied magnetic field (Hall probe measurement). The local modification of the saturation magnetisation ($M_{\text{S},298\,\text{K}} \approx 140\,\text{kA/m}$ at room temperature) is included via the temperature distribution $T(x, y)$ [30]:

$$M_{\text{S}}(x, y) \approx M_{\text{S},298\,\text{K}} - 313 \frac{\text{A}}{\text{K m}} \cdot [T(x, y) - 298\,\text{K}] \qquad (3)$$

The temperature – and therefore $M_{\text{S}}$ – is approximately uniform over the film thickness. The native crossing point of both dispersion relations for the same $M_{\text{S}}$ and $H_{\text{ext}}$ at $k_x = k_y = 0\,\text{rad/cm}$ is called the ferromagnetic resonance (FMR) frequency:

$$\text{FMR} = \frac{1}{2\pi} \sqrt{\omega_{\text{H}} [\omega_{\text{H}} + \omega_{\text{M}}(T)]} \qquad (4)$$

**Conversion efficiency.** To determine the frequency of the conversion efficiency's maximum of spin waves propagating in magnetisation gradients we choose the following simplified approach (all the given parameters are frequency dependent). $S_{21}$ describes the normalised power which is transmitted between the antennas 1 and 2 ($0 \leq S_{21} \leq 1$). A spin wave propagating between these antennas is damped (parameter $\beta$) and converted (parameter $\gamma$):

$$S_{21} = \delta_{\text{antenna}}^{\text{BVMSW}} \cdot \beta^{\text{BVMSW}} \left( \frac{L}{2} \right) \cdot \gamma \cdot \beta^{\text{MSSW}} \left( \frac{W}{2} \right) \cdot \delta_{\text{antenna}}^{\text{MSSW}} \qquad (5)$$

Here, $\delta_{\text{antenna}}$ describes the excitation / detection efficiency of the regarded antenna. $L$ is the distance from antenna 1 to antenna 3 and $W$ is the waveguide width. Neglecting the microwave absorption of the microstrip antennas, the excitation and detection frequency range at antenna 1 and 2 is given by

$$\delta_{\text{antenna}}^{\text{BVMSW}} = 1 - S_{11} \qquad (@\,M_{\text{S},1})$$

and

$$\delta_{\text{antenna}}^{\text{MSSW}} = 1 - S_{22} \qquad (@\,M_{\text{S},2}).$$

The propagation or damping parameter $\beta$ is determined via ($x$ is the propagation distance):

$$\beta(x) = e^{-2 \frac{x}{l_{\text{prop}}}}$$

The propagation length $l_{\text{prop}}$ can be calculated using the group velocity $v_{\text{gr}} = 2\pi \frac{\partial f}{\partial k}$ and the lifetime $\tau$ of the particular spin-wave mode:

$$l_{\text{prop}} = v_{\text{gr}} \cdot \tau$$

Here,

$$\tau = \frac{1}{P_{\text{A}} \alpha \omega} \,.$$

$\alpha = 0.5 \cdot 10^{-4}$ is the Gilbert damping constant, $\omega$ the angular frequency, and $P_A = \sqrt{1 + \left(\frac{\gamma \mu_0 M_S(T)}{4\pi f}\right)^2}$ the ellipticity correction factor [41]. Finally, the conversion spectrum $\gamma$ is given by equation (5):

$$\gamma = \frac{S_{21}}{\delta_{\text{antenna}}^{\text{BVMSW}} \cdot \beta^{\text{BVMSW}}\left(\frac{L}{2}\right) \cdot \beta^{\text{MSSW}}\left(\frac{W}{2}\right) \cdot \delta_{\text{antenna}}^{\text{MSSW}}} \quad (6)$$

**Isofrequency curves.** The isofrequency curves shown in figure 4 are calculated numerically using the equation [35]:

$$(\mu + 1)k_x^2 + (\mu^2 - \nu^2 + 1)k_y^2 + 2\mu\sqrt{\left(-\frac{k_x^2}{\mu} - k_y^2\right)\left(k_x^2 + k_y^2\right)} \cot\left(d\sqrt{-\frac{k_x^2}{\mu} - k_y^2}\right) = 0 \quad (7)$$

The parameters $\mu$ and $\nu$ are given by:

$$\mu = 1 + \frac{\omega_M(T)\,\omega_H}{\omega_H^2 - \omega^2}$$

$$\nu = \frac{\omega_M(T)\,\omega}{\omega_H^2 - \omega^2}$$

Equation (7) connects the two components $k_x$ and $k_y$ with each other. Furthermore, the cotangent was substituted with a Laurent series:

$$\cot x = \sum_{n=0}^{\infty} (-1)^n \frac{2^{2n}\, B_{2n}}{(2n)!} x^{2n-1}$$

We took the first ten terms of the series to approximate the fundamental thickness mode. The $B_{2n}$ are the Bernoulli numbers.

**Simulation.** The micromagnetic simulations are performed using the open-source software mumax³ (v3.9.1) [42]. The influence of the temperature on the saturation magnetisation was implemented via equation (3). The respective dimensions in the $x$, $y$, and $z$ directions are 10 mm x 3 mm x 6.6 µm (1024 x 512 x 1 cells). So, the thickness of the whole spin-wave waveguide is modelled by one cell. The achieved cell sizes are 9.8 µm x 5.9 µm x 6.6 µm. The excitation of the spin waves is done using the Oersted field of a microstrip antenna (50 µm width). The RF current was chosen to be 2 mA to avoid nonlinearities. The excitation signal was applied at the centre of the waveguide (at $x$ = 5 mm). In figure 4, extracts of the propagating wave for $x >$ 5 mm are shown (4 mm x 3 mm, 120 ns after the excitation, stationary regime). To avoid reflections at the edges of YIG (at $x$ = 0 and 10 mm), a 1 mm wide absorbing region was used. Here, the Gilbert damping constant $\alpha$ smoothly and linearly increases from $10^{-4}$ to $10^{-1}$.

**Acknowledgments**

Financial support by the DFG transregional collaborative research center (SFB/TRR) 173 "Spin+X – Spin in its collective environment" (project B04) is gratefully acknowledged.

**Author contribution statement**

M.V. and G.v.F. devised and planned the project. R.A. started the first experiments during his bachelor thesis. M.V. realised the experimental setup, performed the final measurements, analysed the data, performed the simulations, and wrote a first draft of the manuscript. P.P. supported the simulations. A.V.C., B.H. and G.v.F. led the project. All authors discussed the results and contributed to writing the manuscript.

**Competing financial interests**

The authors declare no competing financial interests.

# Adiabatic Control of Spin-Wave Propagation using Magnetisation Gradients

## – Supplementary Materials –

Marc Vogel[1,*], Rick Aßmann[1], Andrii V. Chumak[1], Philipp Pirro[1], Burkard Hillebrands[1] & Georg von Freymann[1,2]

[1] Department of Physics and State Research Center OPTIMAS, University of Kaiserslautern, Erwin-Schroedinger-Str. 56, 67663 Kaiserslautern, Germany

[2] Fraunhofer-Institute for Industrial Mathematics ITWM, Fraunhofer-Platz 1, 67663 Kaiserslautern, Germany

[*] Correspondence and requests for materials should be addressed to M. Vogel (email: mvogel@physik.uni-kl.de)

### Refraction in a gradient of the saturation magnetisation.

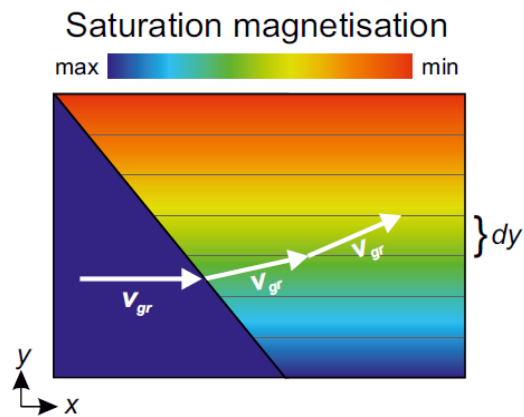

**Figure S1 |** A simple model illustrates the mode conversion process. The saturation magnetisation gradient can be modelled by infinitely thin slices (thickness: $dy$) parallel to the $x$ direction. The spin wave is refracted into the $y$ direction at each slice.

The change of $\vec{k}$ in the $y$ direction can be qualitatively understood using the simple model shown in Fig. S1. Next after entering the gradient region, the direction of the group velocity changes depending on the orientation of the interface. As a consequence, the wave propagates into regions with lower saturation magnetisation. The tangential component of $\vec{k}$ – here it is $k_x$ – is conserved at the refraction at an interface of the shown slices. So, only the $k_y$ component can change. How strong this change $\Delta k_y$ will be is determined via the manifold of possible solutions given by the dispersion relations or isofrequency curves, respectively.

***Schematic experimental setup and sample design.***

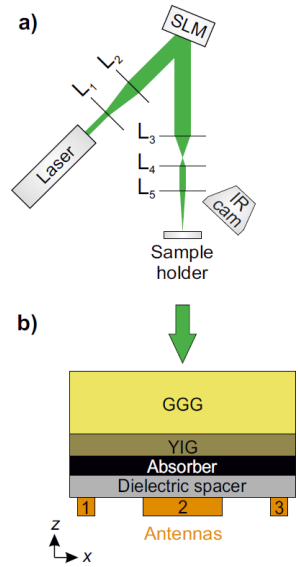

**Figure S2 |** Scheme of the optical experimental setup **(a)** and the sample consisting of a multilayer system **(b)**.

In contrast to the experimental setup in reference [30] in the main text, no acousto optical modulator was used. A green laser creates temperature or respectively magnetisation gradients (see Fig. S2a): a spatial light modulator changes the local phase fronts of the incoming laser beam to create arbitrary intensity distributions / to reconstruct holograms on the sample which heats up locally. An infrared camera measures the resulting temperature distribution.

The sample used in the experiment is schematically shown in Fig. S1b. It consists of a multilayer system (GGG/YIG/absorber/spacer). The laser light impinges from the GGG (Gadolinium Gallium Garnet) side and is absorbed in YIG (Yttrium Iron Garnet) and the black absorber. The dielectric spacer separates the antennas from the sample in order to minimise the thermal contact.

***Additional homogenous light distributions and corresponding local saturation magnetisation.*** In our experiments, we also investigated homogenous light distributions. Let us discuss three cases to create the magnetisation landscapes (see Fig. S3): a rectangle, a triangle, and a "triangle at the bottom" (meaning that the horizontal edge of the triangle points away from antenna 2). In all the cases presented in Fig. S3 the sample is heated via a uniform intensity distribution. But, even a uniform intensity distribution will create a non-uniform temperature profile – a temperature gradient $\vec{\nabla}T$ and, thus, a magnetisation gradient $\vec{\nabla}M_S(T)$ – due to the intrinsic thermal conductivity of YIG. However, the saturation magnetisation $M_{S,2}(T)$ at antenna 2 is kept constant via adjusting the hologram laser power $P_{\text{holo}}$ appropriately.

This is done to compare the cases described in Fig. S3. Antenna 1 and antenna 3 are kept far away from the heated area, thus $M_{S,1}$ and $M_{S,3}$ are almost equal and correspond approximately to the value at room temperature.

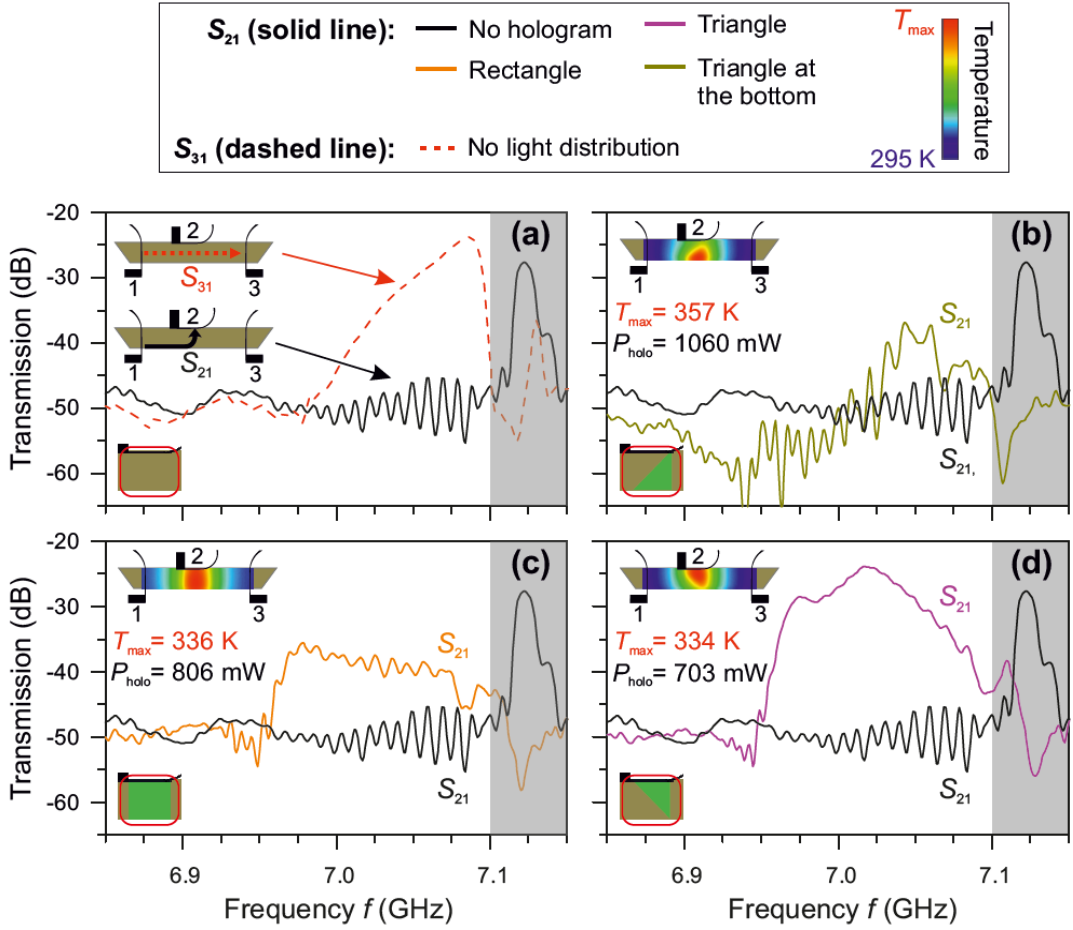

**Figure S3 |** The transmission characteristics for different light intensity distributions (**a**: no illumination, **b**: triangle at the bottom, **c**: rectangle, **d**: triangle) are shown versus the spin-wave frequency. The light patterns are applied in the centre between antenna 1 and 3 – at the position of antenna 2 ("illumination area" in Fig. 1c) – and the corresponding spin-wave transmission ($S_{21}$ and $S_{31}$) is measured. The $S_{21}$ ($S_{31}$) parameter is shown as straight (dotted) line and depicts the transmission of spin waves form antenna 1 to 2 (1 to 3). The light-induced temperature distributions are depicted as colour code (insets, blue: cold, red: hot). $T_{max}$ is the maximal temperature of the colour scale and corresponds to the individual laser power $P_{holo}$. The grey area (7.10 – 7.15 GHz) is above the ferromagnetic resonance frequency for an unheated sample (see the $S_{31}$ parameter) and is not of interest in this work.

For the "triangle at the bottom", one clearly observes a weak $S_{21}$ transmission in the frequency range from 7.025 to 7.075 GHz. However, magnetostatic spin waves are strongly anisotropic and a change in the propagation direction should occur simultaneously with a change in the type of spin-wave mode. If we rotate the triangle counter clockwise by 90 degrees (triangle case) the detected signal at antenna 2 increases drastically. Even the rectangle shows a weak mode conversion. In the latter case, no spin-wave mode conversion is expected since the symmetry of the system is not broken by the laser intensity distribution in the $y$ direction. But regarding the temperature distribution (Fig. S4), the temperature drops down because of von Neumann boundary conditions at the lower edge of the waveguide (at $y$=0.0 mm). Thus, the symmetry in the saturation magnetisation is broken nonetheless. The spin-wave conversion efficiency can be increased further by directly shaping the gradient – as shown in the main text. This is easily realised in our experimental setup by modifying the hologram [30].

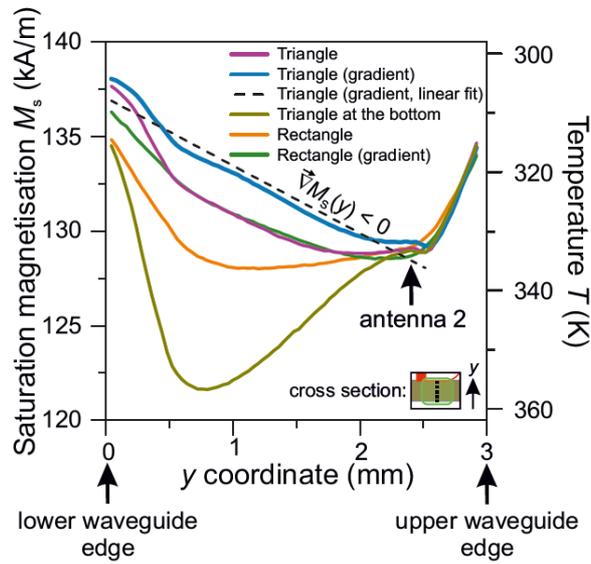

**Figure S4 |** The profiles of the saturation magnetisation along the $y$ coordinate (over the waveguide width) at the centre of the illumination area (see inset, black dotted line) are shown for the different light distributions and laser powers used in Fig. S3. Gradients in the saturation magnetisation with a negative slope cause the mode conversion. In Fig. 2c only the cases with an intensity gradient are presented.

*Micromagnetic simulations.*

1) **Dispersion relations for the saturation magnetisation gradient used in the numeric calculations.** The $M_S$ variation/distribution reaches from about 123 kA/m (maximal heating in the experiment) to 141 kA/m (at room temperature). Thus, the mode conversion area is enlarged compared to Fig. 1b. Mode conversion is possible for an excitation frequency of 7.0 GHz, which is chosen exemplarily in the micromagnetic simulations in Fig. 4. The green arrow depicts the wavevector's transformation inside the gradient area accordingly to the isofrequency curves in Fig. 4a.

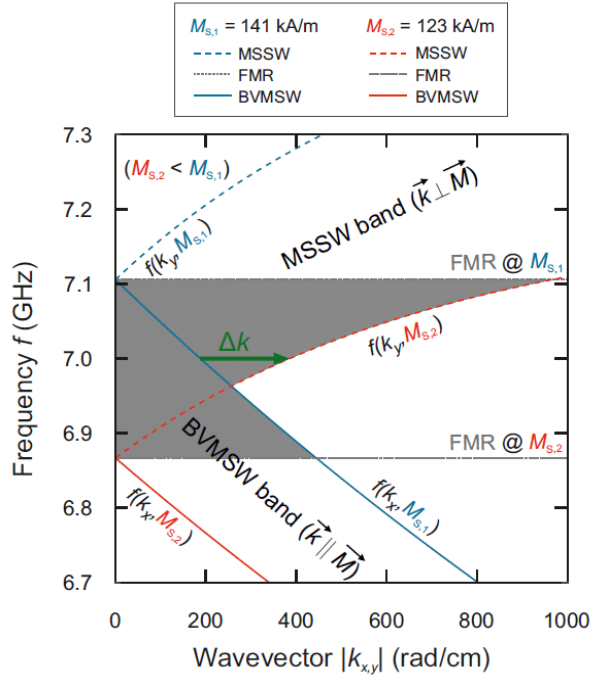

**Figure S5 |** The dispersion relations of BVMSWs and MSSWs are shown for different values of the saturation magnetisation corresponding to the micromagnetic simulations shown in Fig. 4. The change in the wavevector's magnitude due to the conversion process is depicted as green arrow.

2) **Micromagnetic simulations for pure BVMSW behaviour.** If the spin waves are excited at a frequency below the ferromagnetic resonance frequency for the lowest saturation magnetisation, pure BVMSW behaviour is expected. Exemplarily, for $f=6.8$ GHz MSSWs cannot be excited for the minimal saturation magnetisation of around 123 kA/m since the decrease in $M_S$ is too small (see the dispersion relations in Fig. S5). Consequently, no isofrequency curves for MSSWs are occurring above the line corresponding to the critical angle $\varphi_c$. The spin-wave propagation in the gradient area is the same as in Fig. 4 for waves entering the $\vec{\nabla}M_S$ region below $M_{S,\text{FMR}}$. In the case of a rectangular magnetisation gradient area (Fig. S6a), the spin waves propagating at $M_S \approx 141\,\text{kA/m}$ enter the gradient at a certain value $M_{S,\text{in}}$ defined by the coordinate in the direction perpendicular to the propagation direction. Thus, the spin-wave wavevector, and, respectively, the wavelength change differently in different positions over the waveguide's width resulting in a bending of the phase fronts and in an adjustment of the group velocity's direction. The corresponding isofrequency curves illustrate how this process is happening: the spin waves at $M_S \approx 141\,\text{kA/m}$ propagate into the gradient region at $M_{S,\text{in}}$. The tangential component of $\vec{k}$ (with respect to the interface) is conserved and only $k_x$ is changed. After entering $\vec{\nabla}M_S(x,y)$, $k_x$ is conserved and only $k_y$ is modified since the translational symmetry of the system is broken due to the magnetisation gradient in the $y$ direction. The change $\Delta k_y$ is due to quasi-adiabatic refraction of the spin waves in the gradient area, which is much larger than the wavelength [29]. As a consequence, $\vec{k}$ rotates in the $x$-$y$-plane. In the triangular case (Fig. S6b) additional refraction at the interface to the gradient area occurs, which changes the $x$ and $y$ component of the wavevector.

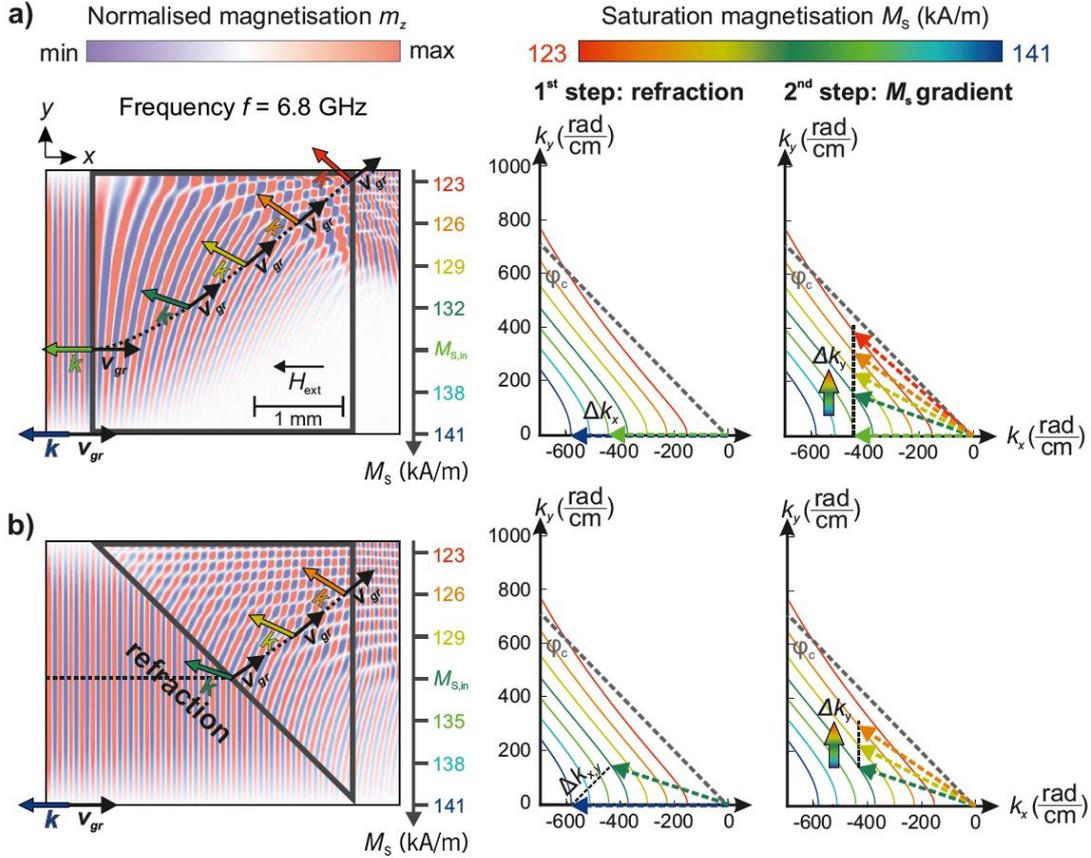

**Figure S6 |** Reference simulations for $f$ =6.8 GHz (pure BVMSW behaviour) in addition to Fig. 4 in the main text.

3) ***Micromagnetic simulations – no upper waveguide edge / energy flow.*** In this case, the waveguide is enlarged at the upper edge. Outside of the black area (gradient area), the saturation magnetisation $M_S$ is kept constant (~141 kA/m in the left region, ~123 kA/m in the top region). Since no reflections are occurring at the top, the phase fronts can be seen clearer (no superposition of incoming and reflected waves). Furthermore, a fast Fourier transform (FFT) in time shows how the energy flows through the gradient region. On the left side of the gradient region a standing wave is formed. For a frequency $f$ =7.0 GHz, the triangular hologram is much more effective compared to a rectangular one.

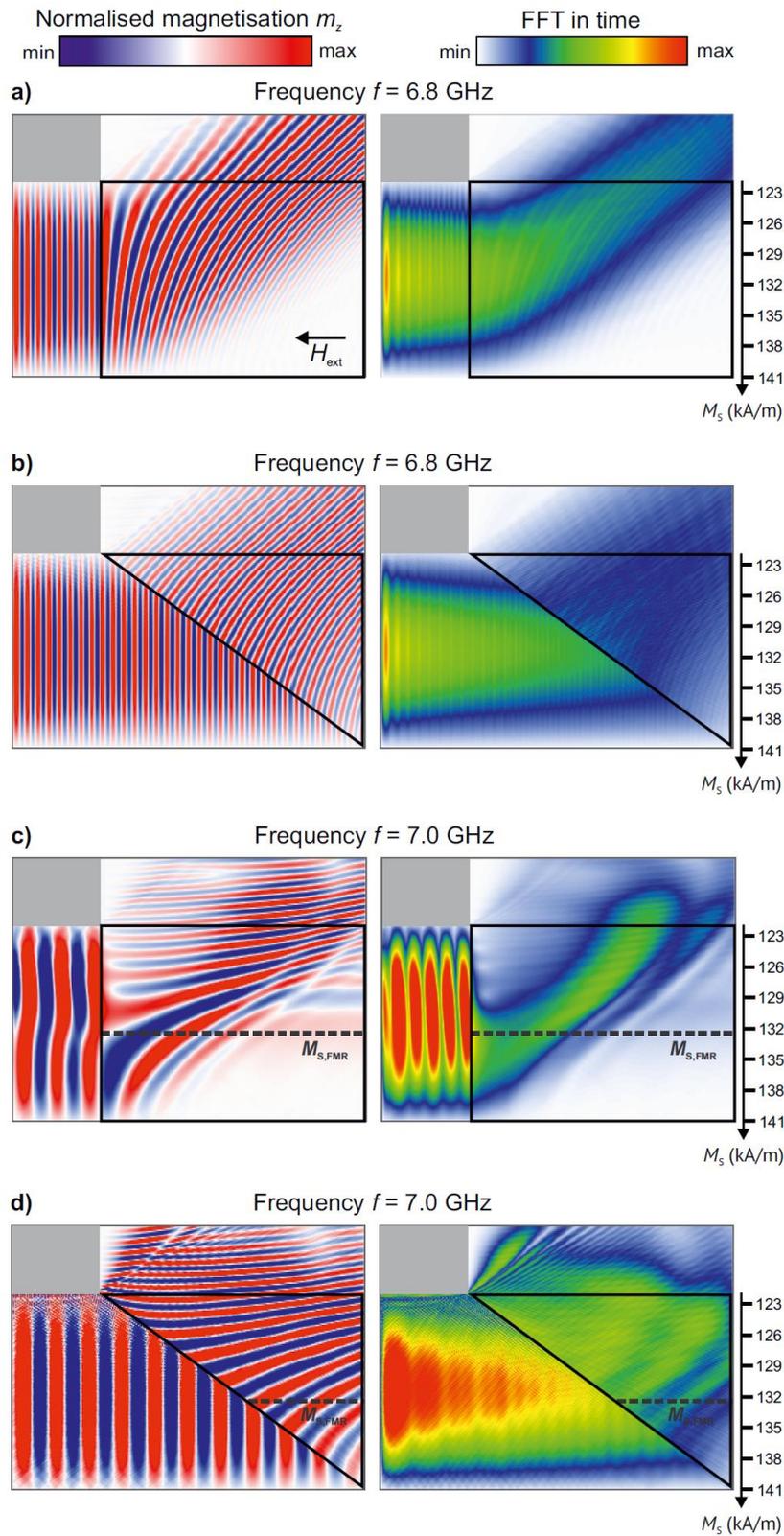

**Figure S7 |** Micromagnetic simulations (left column) for the frequencies 6.8 GHz (**a**, **b**) and 7.0 GHz (**c**, **d**) and different magnetisation landscapes (rectangle: **a**, **c** & triangle: **b**, **d**). The energy flow is shown as fast Fourier transformation in time in the right column. The $M_S$ gradient is the same as in Fig. 4.

***The critical angle and its dependency on the saturation magnetisation.*** The critical angle $\varphi_c$ defines if MSSW modes can exist for the propagation into a certain direction $\varphi$. More information concerning $\varphi_c$ can be found in e.g. reference [36]. It is defined as:

$$\varphi_c(T) = \arctan\left(\sqrt{\frac{H_{\text{ext}}}{M_S(T)}}\right)$$

The saturation magnetisation dependency of the critical angle is shown in the following plot for $H_{\text{ext}} = 143$ kA/m (equals $\mu_0 H_{\text{ext}} \approx 180$ mT):

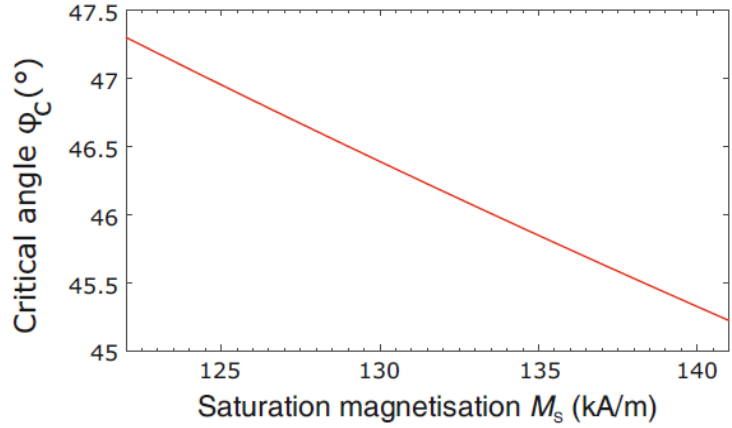

**Figure S8 |** The critical angle as a function of the saturation magnetisation.

The average value over the shown range of $M_S$ is approximately 46 degrees.

***Additional movies.*** In Fig. 4 and S6, the numerical calculations of the spin-wave propagation in different gradients of the saturation magnetisation is shown 120 ns after the excitation. These additional movies show the full micromagnetic simulations for 0 – 120 ns. The time step between every frame of the movie is 1 ns. The spin-wave is excited in the centre of the waveguide. The landscape of the saturation magnetisation is placed on the right side of the excitation area. On the left side, the uninfluenced spin wave is shown as a reference.